# Chiral quantum mechanics (CQM) for antihydrogen systems

G. Van Hooydonk, Ghent University, Faculty of Sciences, Krijgslaan 281, B-9000 Ghent (Belgium)

Abstract. A first deception of QM on $\underline{H}$ already appears in one-center integrals for two-center systems [G. Van Hooydonk, physics/0511115]. In reality, full QM is a theory for chiral systems but the QM establishment was wrong footed with a permutation of reference frames. With chiral quantum mechanics (CQM), *the theoretical ban on natural $\underline{H}$ must be lifted as soon as possible.*
Pacs: 34.10.+x, 34.90.+q, 36.1.-k

For molecular cations $HH^+$ and $\underline{H}H^+$, we proved recently that QM is inconsistent on antihydrogen by a misjudgment on the *proton-antiproton attraction*, which resides primarily in one-center integrals [1]. In ab initio calculations for this molecular cation [2], the secular equation

$$\begin{vmatrix} H_{AA} - W & H_{AB} - SW \\ H_{BA} - SW & H_{BB} - W \end{vmatrix} = 0 \qquad (1)$$

gives symmetric and antisymmetric solutions

$$W_S = (H_{AA} + H_{AB})/(1+S) \qquad (2a)$$
$$W_A = (H_{AA} - H_{AB})/(1-S) \qquad (2b)$$

Positional equivalence secures that $H_{AA}=H_{BB}$ but this *seemingly straightforward* conclusion can be deceptive. Mathematically, their equality is proved with

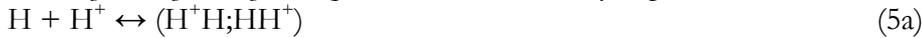
$$H_{AA} = \int \psi_A (W_H - e^2/r_B + e^2/r_{AB}) \psi_A d\tau = W_H + e^2/r_{AB} + J \qquad (3a)$$
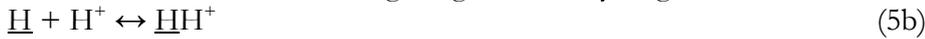
$$H_{BB} = \int \psi_B (W_H - e^2/r_A + e^2/r_{AB}) \psi_B d\tau = W_H + e^2/r_{AB} + J \qquad (3b)$$

giving identical results for orbitals, centered at either nucleus since the same Hamiltonian is used for molecular cations $HH^+$ and $H^+H$. Hence, the identity

$$H_{AA} = H_{BB} \qquad (4)$$

*can only be obtained by inverting (a permutation of) the reference frame for the two centers of the species, although a permutation always has symmetry-consequences.* With chemistry, a permutation for reversible reaction (5)

$$H + H^+ \leftrightarrow (H^+H; HH^+) \qquad (5a)$$

refers to *a proton perturbing a fixed atom* and to *an atom perturbing a fixed proton.* The Hamiltonian for hydrogenic cation $HH^+$ however changes sign for antihydrogenic cation $\underline{H}H^+$ obeying reaction

$$\underline{H} + H^+ \leftrightarrow \underline{H}H^+ \qquad (5b)$$

Equivalent reactions (5a) are described with equivalent reference frames, *one subject to a permutation* but, in reality, only one reference frame is allowed. If a parity operator is needed, this is readily provided with (5b). These elementary facts lead to a *misconception about proton-antiproton attraction*, if the QM machinery were not checked for permutations (5a), i.e. for chiral behavior (5b).

(i) Scaling $r_{AB}$ with $Da_0$, gives scaled Coulomb interactions $\pm 1/D$ and can be used for a graphical justification for (4). The triangular $1/D$ graphs in Fig. 1a and 1b have advantages to become clear below. Reminding (5), $H^+$ starts perturbing atom H at 0,0 (in a D-graph, this is at infinite separation $0,\infty$). Fig. 1a and 1b illustrate a change of subscripts A and B in (3) with the notation $AB^+$ for $HH^+$ (5). Fig. 1a for system $B^+...A$ in (5) gives (3a), reproduced identically in Fig. 1b for $A^+...B$ in (5) and (3b). Identical triangles *prove* the validity of QM result (4).
But at this stage, convention enters the scene. To bridge the gap of one half Hartree between the cation and the atom, two options are allowed: going from atom to cation, repulsion or $+1/D$ can be used but then going from cation to atom requires attraction or $-1/D$. In QM, the leading terms in $H_{AA}$ and $H_{BB}$ (3) are $W_H + e^2/r_{AB}$ [1,2], where $W_H = -\tfrac{1}{2}e^2/a_0$. Anyone would conclude at this stage that the energy of the molecular hydrogen cation $HH^+$ or $H^+H$ is based exclusively on *classical Coulomb proton-proton repulsion* $+e^2/r_{AB}$. This is the long-standing premise of QM that $\underline{H}H^+$ and $H^+\underline{H}$ are *exotic* systems and must be forbidden in nature, due to their *exotic proton-antiproton attraction* $-e^2/r_{AB}$ [1]. This attraction contradicts standard analytical QM result (3), e.g. its unavoidable Coulomb repulsion $+e^2/r_{AB}$ [1]. However, the Coulomb duality above remains and is easily proved by scaling (3). Standard scaling with $+\tfrac{1}{2}e^2/a_0$ gives $-1+2/D$ of *repulsive Coulomb type* for QM result (3) but equally valid scaling with $-\tfrac{1}{2}e^2/a_0$ would give $+1-2/D$ of *attractive Coulomb* *type, characteristic for (5b). Choosing the repulsive option and rejecting or forbidding the attractive one, as suggested by QM, seems plausible but must not be absolutely valid.* With (3), QM succeeded in wrong footing most of us on the character of the nucleon interaction, which led to the veto for natural H̲ and its systems [3]. This *seemingly obvious but certainly not absolutely valid* conclusion finds its origin in the permutation of the reference frame as in (3a) and (3b), as illustrated in Fig. 1a and 1b and as proved with scaling. A permutation of a reference frame must always have consequences *for symmetry*. For an x-axis, a permutation would transform the +x semi-axis in the –x semi-axis, just like the parity operator appearing when bridging the energy gap with repulsion +1/D and attraction –1/D. *This ambiguity with Coulomb's law, just like with the permutation, must be removed.* If QM deals correctly with a permutation of reference frames, *it must also use rather than forbid Coulomb attraction –1/D, typical for (5b)* [1].

(ii) This brings us to the graph with a single reference frame in Fig. 2 *without permutation*. All interactions, including that of the nucleons, needed to bridge gap 0,-1 must have switched sign with a displacement of the lepton from one to the other center, *while remaining in the same reference frame for the nucleons*. The surfaces of the triangles in Fig. 1a and Fig. 1b are exactly equal to these of Fig. 2. But the 2 triangles in Fig. 2 are *complementary*: they are *mutually exclusive left- and right-handed triangles*. With Lord Kelvin's definition of chiral behavior [4], *these two triangles can never be made to coincide without leaving the plane*. Hence, the left-right difference in Fig. 2 gives away a discrete transition from repulsive to attractive forces (or vice versa), *seemingly not acknowledged for by* (4) or by Fig. 1a and 1b, *where it is disguised by a permutation of the reference frame*. If in Fig. 2, the left-handed triangle for B$^+$...A (5) stands for repulsive interaction +1/D, as in QM with the *proton-proton repulsion* +$e^2/r_{AB}$, then the right-handed triangle for B...A$^+$, also in (5), can only stand for attractive –1/D or for *proton-antiproton attraction* –$e^2/r_{AB}$, forbidden by the *false man-made QM premises* and by conventional physics at large [3]. This proves the reality of reaction (5b) even for QM, since pair H;H̲ is chiral, *due to a permutation of charges*, just like cation pair HH$^+$; H̲H$^+$ [1].

*The rigor of QM does not allow fooling with permutations.* In fact [1], the *forbidden* Coulomb *proton-antiproton attraction* appears already at the level of the two one-center integrals J, defined as

$$J = \int \psi_A(-e^2/r_B)\psi_A d\tau \equiv \int \psi_B(-e^2/r_A)\psi_B d\tau = -e^2/r_{AB} + \text{other repulsive terms...} \quad (6)$$

Although only repulsive proton-proton Coulomb term +$e^2/r_{AB}$ appears explicitly in (3), full QM (6) rightly shows that the leading term in J contains the *so-called forbidden Coulomb proton-antiproton attraction* –$e^2/r_{AB}$ as expected from the left-right symmetry in Fig. 2. With intersecting triangles, a Coulomb cusp is generated when repulsive –1+1/D and attractive –1/D interactions are in equilibrium, which is impossible with Fig. 1. Translation in a fixed reference frame may leave the laws of physics intact; a permutation of an axis creates a parity operator. In higher order, QM also deals with two-center integrals but the necessary algebraic Coulomb switch already appears at the level of one-center integrals as in Fig. 2. This Coulomb switch plays a dominant role when (6) is introduced in (3) but, due to the symmetry of the primary Coulomb forces, it is less visible in the observed PEC (potential energy curve) of the molecular cation [1].

*Full QM proves a perfect theoretical framework to deal with chiral systems and can readily be called chiral quantum mechanics (CQM). In this way, one prevents QM from becoming deceptive on proton-antiproton attraction in natural systems* like the molecular hydrogen cation [1], the neutral hydrogen molecule [3] as well as the atomic species hydrogen [5]. In any case, *the theoretical ban on natural H̲, persistently but wrongly promoted to the status of a QM premise, must be lifted as soon as possible*, even while *wrong artificial H̲ experiments* are in progress, as argued in [6].

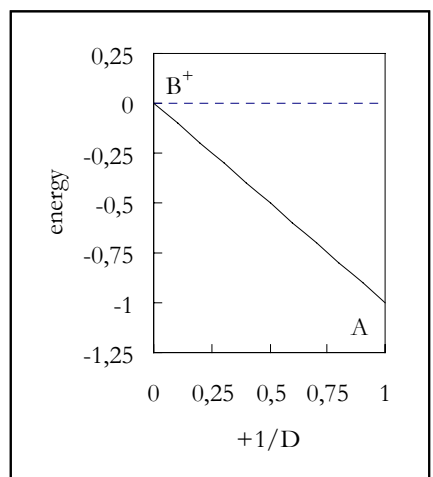

Fig. 1a Energy triangle for perturbing cation B⁺ acting on atom A (fixed)

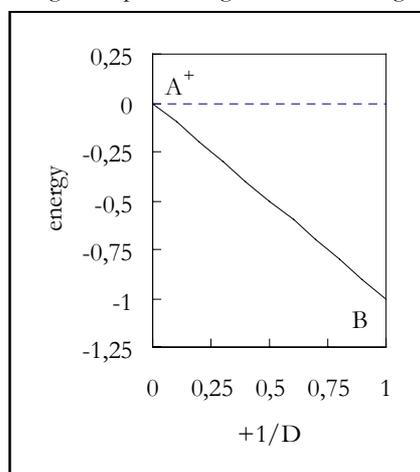

Fig. 1b Energy triangle for perturbing cation A+ acting on atom B (fixed)

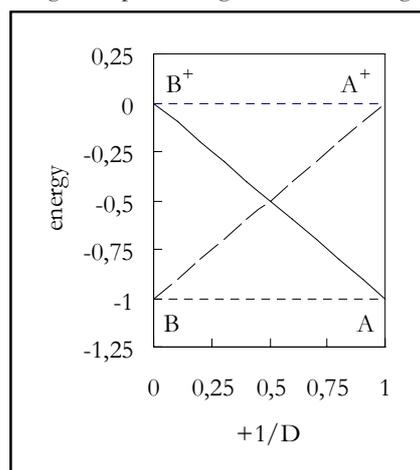

Fig. 2 Left- and right-handed energy triangles for reaction (5), with repulsive –1+1/D as well as unjustly forbidden attractive –1/D